\documentclass[preprint2]{aastex}
\usepackage{spr-astr-addons}
\usepackage{graphicx}
\newcommand{\gtapprox}{\raisebox{-0.5ex}{$\,\stackrel{>}{\scriptstyle
\sim}\,$}}

\begin{document}
\title{Long Gamma-Ray Burst Progenitors: Boundary Conditions and Binary Models
}

\shorttitle{GRB Progenitors}        
\shortauthors{E.P.J. van den Heuvel and  S.-C. Yoon}

\author{ E.P.J.van den Heuvel and
S.-C. Yoon}
\affil{Astronomical Instiutute ``Anton Pannekoek''  \& Center for High Energy Astrophysics, University of Amsterdam, The Netherlands  and Kavli Institute for Theoretical Physics, University of California, Santa Barbara}





\begin{abstract}
The observed association of Long Gamma-Ray Bursts (LGRBs) with peculiar Type Ic supernovae gives support to Woosley`s collapsar/hypernova model, in which the GRB is produced by the collapse of the rapidly rotating core of a massive star to a black hole. The association of LGRBs with small star-forming galaxies suggests low-metallicity to be a condition for a massive star to evolve to the collapsar stage. Both completely-mixed single star models and binary star models are possible. In binary models the progenitor of the GRB is a massive helium star with a close companion. We find that tidal synchronization during core-helium burning is reached on a short timescale (less than a few millennia). However, the strong core-envelope coupling in the subsequent evolutionary stages is likely to rule out helium stars with main-sequence companions as progenitors of hypernovae/GRBs. On the other hand, helium stars in close binaries with a neutron-star or black-hole companion can, despite the strong core-envelope coupling in the post-helium burning phase, retain sufficient core angular momentum to produce a hypernova/GRB.
\end{abstract}



\section{Introduction}
\label{intro}
About a year after the discovery of the first optical afterglow of a Gamma-Ray Burst  (GRB) by van Paradijs et al. (1997), two of van Paradijs' students discovered the first supernova associated with a long-duration GRB: SN 1998bw/GRB980425 (Galama, Vreeswijk et al. 1998). This supernova appeared to be highly peculiar and energetic. It is of class Ic, which means that it has no H or He in its spectrum. Its outflow velocities of  $> 30 000$  km/s were very much larger than the $²10 000$ km/s seen in ``ordinary'' 
Type Ic supernovae and the total kinetic energy in SN1998bw was $> 10^{52}$ ergs: at least an order of magnitude larger than in other supernovae. Theoretical modeling by Iwamoto at al. (1998) showed that the exploding star must have been a Carbon-Oxygen star with a mass in the range 6 to 13 $M_\odot$, which had a collapsing core > 3 $M_\odot$. The latter is too large to leave a neutron star, implying that this was the first-ever observed birth event of a stellar-mass black hole (Iwamoto et al. 1998).
The discovery of SN1998bw was a beautiful confirmation of the ``collapsar'' (``hypernova'') model proposed by Woosley (1993). According to this model the collapse of the rapidly rotating core of a massive star to a black hole will leave behind a rapidly rotating torus of extremely hot nuclear matter around the black hole. Internal friction in this keplerian torus causes its matter to spiral in towards the black hole within a few minutes, generating so much heat in this process that part of the matter is blown away in directions perpendicular to the plane of the torus with relativistic velocities. Woosley speculated that these relativistic ``jets''  of matter might produce a GRB. SN 1998bw appeared to confirm the predictions of Woosley`s ``collapsar''  (``hypernova'') model. 
Although GRB980425 was, as a GRB, intrinsically quite faint and nearby ($z$=0.0085), which at first cast some doubt on the idea that genuine long-duration GRBs would in general be the birth events of stellar black holes, the discovery of the association of the really ``cosmological'' gamma-ray burst GRB 030329 ($z$= 0.17) with a supernova with a spectrum and lightcurve almost identical to those of SN1998bw (e.g. Hjorth et al. 2003) confirmed beyond any reasonable doubt the association of long GRBs (abbreviated further as LGRB) with the death events of very massive stars and the formation of black holes. Indeed, while the lightcurves of the optical transients (OTs) associated with LGRBs are often dominated by the radiation from the relativistic outflow of the GRB, numerous LGRBs have shown late-time ``bumps'' consistent with the presence of underlying supernovae (e.g. Bloom et al. 1999; Galama et al. 1999; Levan et al. 2005).  For a review see Woosley and Bloom (2006).
These discoveries have given strong credence to Woosley`s (1993) model as the ``standard'' model for the production of the LGRBs, and this model has been worked out in more detail by Woosley and collaborators (e.g. MacFadyen and Woosley 1999; Woosley and Heger, 2006). To distinguish these very energetic and peculiar Ic ``supernovae'' associated LGRBs from the more ordinary Ibc supernovae, we will in this paper call them ``hypernovae''. 
In order to finish with a pure CO-core of mass $> 6 M_\odot$, a star must have started out on the main sequence with a mass > 30$M_\odot$, which implies that the LGRBs are associated with the most massive stars. Here we will discuss further evidence linking indeed the LGRBs with such stars, and examine under which circumstances a star could lose its entire H- en He-rich envelope before collapsing to a black hole. It appears that the removal of the envelope by a binary companion might be an attractive possibility.

\begin{figure}[ht!]
\begin{center}
\includegraphics[width=0.48\textwidth]{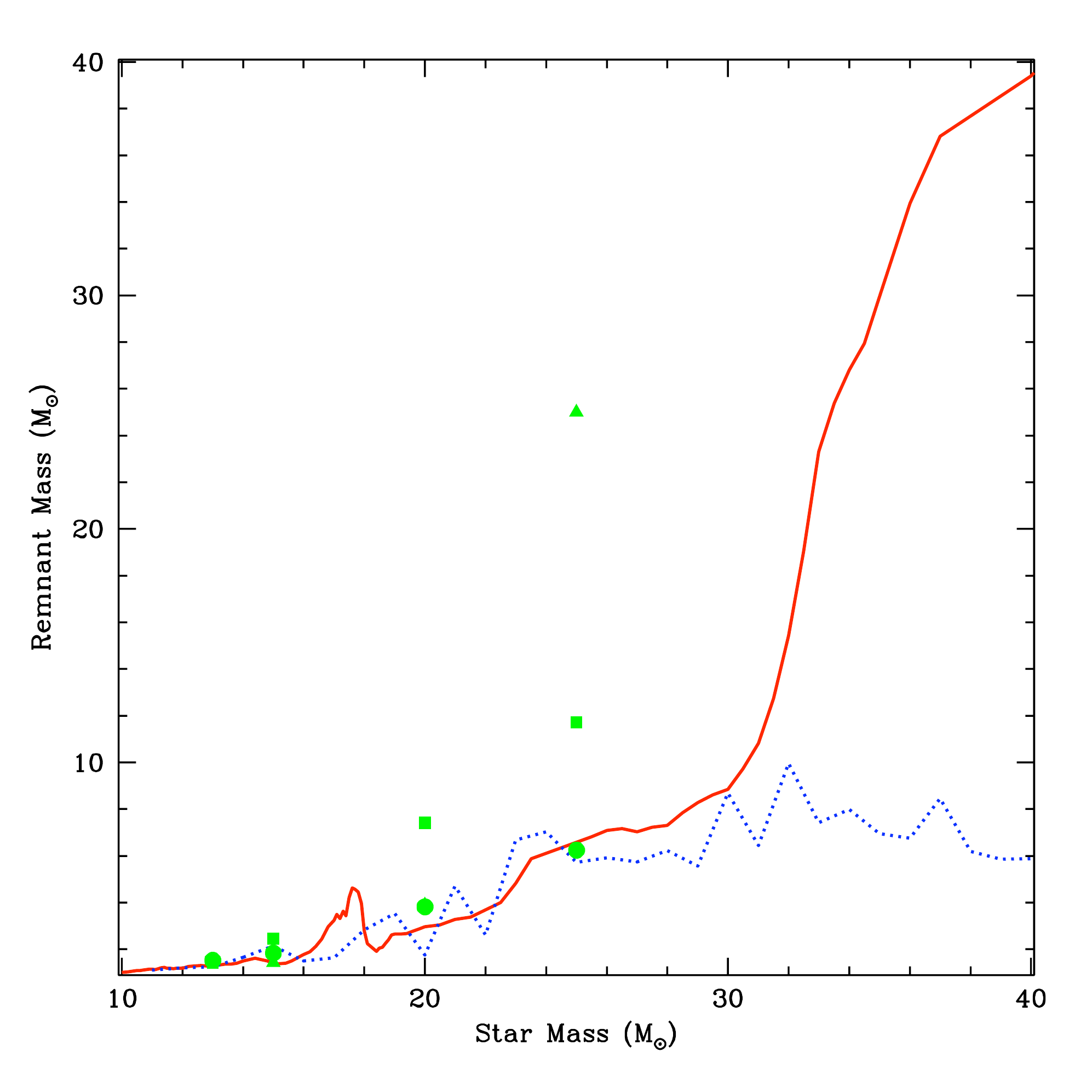}
\end{center}
\caption{Mass of collapsed remnant as a function of initial main-sequence progenitor mass from the analysis by Fryer (2006), for both the Limongi \& Chieffi (2006) and Woosley et al. (2002) stellar progenitors. The lines are derived from the Woosley et al. (2002) progenitors: dotted line refers to solar metallicity, solid line refers to very low metallicity. The points are derived from the Limogni and Chieffi (2006) models: circle -solar, square 0.2 solar, triangle - zero, metallicities. Around 20 solar masses the outcome depends sensitively on the stellar evolution code used. Credit: C.L.Fryer (2006) }
\label{fig:1}       
\end{figure}

\section{Host Galaxy Characteristics: further evidence for an association of the LGRBs with the most massive stars.
}
\label{sec:2}

In a very important recent paper, Fruchter et al. (2006) reported that the environments of LGRBs are strikingly different from those of the ``ordinary''  core collapse supernovae of types Ib,c and II.
Using Hubble Space Telescope imaging of the host galaxies of LGRBs and core-collapse supenovae they found that the GRB are far more concentrated on the very brightest regions of their host galaxies than are the supernovae. Furthermore, they found that the host galaxies of the GRBs are significantly fainter and more irregular than the hosts of the supernovae.
Theoretical work (Fryer, 2004, 2006) shows that stars which started out on the main sequence with masses between 8 and 20 $M_\odot$ leave neutron stars as remnants, while the cores of stars more massive than about 20 $M_\odot$ collapse to black holes. Figure 1, after Fryer (2006) shows that this happens irrespective of initial metallicity, although the black holes produced at lower metallicity tend to be much more massive than those from higher metallicity stars. In view of the slope of the IMF, some 75 per cent of the deaths of stars >8 $M_\odot$ arise from the mass range 8-20$M_\odot$, and only some 25 per cent from masses > 20$M_\odot$. Therefore, the bulk of the core collapse supernovae will be neutron-star forming events. It thus appears that the neutron-star forming events follow the normal light distribution of their host galaxies, whereas the LGRBs are concentrated strongly on the brightest parts of these galaxies. Another striking difference is that while half of the hosts of the ``normal'' core collapse supernovae are Grand Design (GD) spiral galaxies, only one out of the 42 hosts of the LGRBs is a GD spiral, the other 41 being smaller and more irregular galaxies. [In the case of the one GD spiral it is still very well possible that the real host is a small SMC- or LMC-like satellite of this spiral galaxy, which at this distance cannot be separately recognized].

The brightest patches of the irregular and small host galaxies of LGRBs are ``clumps'' of massive stars. This follows from the fact that these hosts are generally found to be very blue ( Fruchter et al. 1999; Sokolov et al. 2001) and have strong emission lines (Bloom et al. 1998; Vreeswijk et al. 2001), suggesting a significant abundance of young massive stars. At the large redshifts of the GRB hosts it is impossible to distinguish the stellar content of the bright emission line spots (the entire HST image of a host is often smaller than an arcsec), but nearby small irregular starforming (``starburst'') galaxies serve as a good example of what is going on in these small GRB hosts. A nearby example of such a galaxy is NGC 3125 which was studied by Hadfield and Crowther (2006). These authors find that the bright spots of this galaxy consist of large concentrations of O- and WR-type stars, which number of order 10 000 in this galaxy. The galaxy has a metallicity like that of the SMC/LMC (between 0.2 and 0.5 solar) and its brightest clump has at least four dense star clusters of > 200 000 solar masses, each with some 600 O-stars. A few of the hosts of relatively nearby LGRBs associated with hypernovae show similar characteristics. The host of SN1998bw is an LMC-size star-forming galaxy; the host of GRB060218 is SMC size; the host of GRB030329 is a z=0.17 undetectable, indicating that its size must be smaller than that of the SMC, and the host of GRB970228 at z=0.67 is not larger than the LMC. 

Recently Wolf and Podsiadlowski (2006), statistically studying part of the host galaxy sample of Fruchter et al. (2006), concluded that the typical LGRB host galaxy is of LMC size. They found, on the basis of the metallicity-luminosity relation for star-forming galaxies, that LGRB models that require a sharp metallicity cut-off below 0.5 solar metallicity are effectively ruled out as they would require fainter host galaxies than are observed. They therefore conclude that metallicities up to 0.5 solar must be allowed by models for LBRBs/hypernovae. As, however, in these irregular galaxies the metallicity may vary wildly from place to place, it is not clear to us whether not the LGRBs might arise from areas in the hosts of much lower metallicity, while the average metallicity of the host might still be up to of order 0.5 solar.


\section{Possible reasons why small  ``starburst-like'' galaxies are the prime sources of LGRBs}
\label{sec:3}

These reasons can be divided into two broad categories:
(1)	Metallicity-related,
(2)	Starburst-related.

As to Category (1): the wind mass-loss rates from massive stars are known to be metallicity-related: Mokiem at al.(2006) find from observations of O- and B-supergiants in the Local Group galaxies that he wind mass-loss rates scale roughly as $\dot{M}_w \propto Z^{0.78}$, where $Z$  is the abundance of the elements heavier than helium. This implies that at lower metallicities, such as in the SMC and LMC (0.2 and 0.5 solar, respectively) massive stars lose (much) less mass during their evolution than in our galaxy. Therefore, they are more likely to finish as a black hole. Indeed, one observes that in the LMC half of the four known persistent High Mass X-ray Binaries (HMXB) harbour a black hole while in our Galaxy only one out of the over 20 known persistent HMXBs harbours a black hole (Cygnus X-1). It thus appears that at low Z, black-hole production is more efficient. In addition, a requirement for producing a ``hypernova'' is that at the time of the core collapse, the star is still rotating sufficiently rapidly to enable the formation of a disk or torus around the black hole (MacFadyen and Woosley 1999). Lower wind mass-loss rates imply also lower angular momentum loss rates, which will increase the probability of having still a sufficiently rapidly rotating stellar core at the time of the collapse.

As to Category (2): It is well-known that during a starburst massive dense star clusters form with many hundreds, if not thousands, of massive OB stars. For example, many such massive young globular clusters are observed in the pair of Antennae Galaxies.  In massive young globular clusters a variety of dynamical interactions take place between massive stars, massive binaries and stellar remnants (black holes, neutron stars) ranging from direct collisions to companion exchanges in binary systems, and to the formation of so-called Intermediate Mass Black Holes (IMBHs) with masses of order 100 to 1000 solar masses (Portegies Zwart et al., 2002, 2004, 2006).
These can be unique events, which do not occur in any other stellar environment. Kulkarni (2006) suggested that LGRBs might be related to such unique events that can occur only in starburst galaxies.  This interesting idea merits to be further worked out, but at present not much further can be said about it.
For this reason we will here only concentrate on the possible relation between LGRBs and metallicity.
In order to make a hypernova such as the ones observed to coincide with the LGRBs, the two following conditions should be fulfilled: \\
(1)	the star must have lost its H- and He-rich outer layers;\\
(2)	At the time of core collapse, the core should have specific angular momentum in the range 
\begin{equation}
\label{eq:1}      
            J (CO-core) = (3-20) \times 10^{16}    \mathrm{[cm^2/s]}                                                     
\end{equation}
In order to fulfill these two conditions, two possible scenarios have been proposed:\\
(i)	Completely-mixed single-star evolution of a rapidly-rotating low-metallicity star (Yoon and Langer 2005; Woosley and Heger 2006).\\
(ii)	Binary mass exchange, where the star achieves and maintains its rapid rotation due to tidal synchronization in a close binary (Izzard et al. 2004; Podsiadlowski et al. 2004).\\
We now separately discuss these two possible scenarios.

\section{Completely mixed single-star models of low metallicity}
\label{sec:4}
In this case the rapid rotation of the star keeps it completely mixed by meridional circulation during its entire H-burning evolution. The low metallicity causes the wind mass- and angular-momentum-loss rates to be small such that the star keeps rotating rapidly until the end. The complete mixing makes that by the end of hydrogen burning the star has become a complete helium star (the weak wind has by that time carried off the thin hydrogen envelope that still surrounded the helium core). Yoon and Langer (2005) calculated such an evolution for a star which started out with M= 40 $M_\odot$ and Z= $10^{-5}$ and find that it evolves into a rapidly rotating pure helium star of 32 $M_\odot$, which after 600 years of C-burning undergoes core-collapse to a black hole with sufficient angular momentum to make a hypernova. They find that this type of evolution follows if the star starts out with an equatorial rotation velocity of $³ 0.5$ times the critical one.
Later calculations by these authors suggest that up to  Z =0.2 solar the stars still follow this evolutionary path. Woosley and Heger find that it would still work up to Z = 0.33 solar. For higher Z this single star model no longer works. If the conclusion of Wolf and Podsiadlowski (2006) mentioned in section 2 would strictly hold, i.e. if models should work up to Z =0.5 solar, these single star models would be ruled out. However, as mentioned at the end of section 2, due to the patchy distribution of metallicity in irregular starburst galaxies, there could easily be patches with SMC-like (Z=0.2) metallicities in the irregular hosts and therefore certainly these completely mixed single star models cannot be ruled out. In the calculations of Yoon and Langer (2005) these stars still have a helium-rich envelope, which would lead to a Type Ib supernova, but later calculated models (Yoon, Langer and Norman 2006) and also some of the Woosley and Heger (2006) models lose this envelope by wind such that they would produce a Type Ic supernova. 

\section{Binary Models; can LGRBs be the formation events of Black-Hole X-ray Binaries?}
\label{sec:5}
\subsection{Introduction}
\label{subsec:5.1}
The first ones to consider binary models for making LGRBs were Fryer and Woosley (1998). Their model was, however, not a core-collapse model, but one in which an already existing black hole in an X-ray binary spiraled down into the helium core of its massive companion, as a result of a Common-Envelope phase. Although interesting, we will not consider such models here and only concentrate on ``hypernova'' models in which the LGRB coincides with the core-collapse event in which a black hole is formed.\\  
Izzard et al. (2004) and Podsiadlowski et al. (2004) were the first to consider the role that binary systems might play in producing such ``hypernova'' events.
At present some twenty close X-ray binaries are known that consist of a black hole and a low-mass companion star (see McClintock and Remillard, 2006). The black hole in such systems typically has a mass between 3 and 20 $M_\odot$, and the companion is a Roche-lobe filling star with a mass < 2$M_\odot$. The orbital periods are in general less than a few days, and in many cases less than 0.5 day. In the system of X-ray-Nova Sco 1994 (J1655-40) the F-type companion of the 7 $M_\odot$ black hole has an overabundance of alpha-type elements 
such as S, Mg and Si of more than one order of magnitude (Israelian et al. 1999). This is just what one expects if the outer layers of the star of which the core collapsed to the black hole were ejected in a supernova-like event and polluted the outer layers of the F-type companion. It thus appears that in this black-hole X-ray binary a hypernova-like event took place. Podsiadlowski et al. (2004) propose that in all of these low-mass black hole X-ray binaries the formation event of the black hole produced a LGRB.
The formation of these BH-LMXBs requires a preceding Common-Envelope (CE) phase of an initially wide binary system consisting of the massive progenitor star of the black hole together with a distant low-mass companion star (e.g. see van den Heuvel and Habets 1984; Brown et al. 1996; Nelemans and van den Heuvel, 2001). During this CE phase the low-mass companion spiraled down deeply into the envelope of the massive companion resulting in a very close binary system consisting of the helium core of the massive star together with its low mass-mass main-sequence companion $(< 2M_\odot)$. Izzard et al. (2004) and Podsiadlowski et al. (2004) suggested that tidal forces in this close binary keep the helium star in synchronous (=rapid) rotation, allowing it to have sufficient angular momentum at the time of its core collapse to produce a hypernova. These authors, however, did not calculate the timescales on which tidal synchronization in such binaries can be achieved. In order to see whether such a model can work, one has to calculate these timescales as well as the timescales on which the rotation of the contracting stellar core is synchronized with the outer envelope of the star. These two problems we will consider here.

\subsection{Timescales for synchronization of helium stars in close binaries with a    
      main-sequence companion.}
\label{subsec:5.1}      
 We consider helium stars of 8 and 16 $M_\odot$, which are probably representative for the progenitors of the black holes in LGRBs. Helium-burning helium stars with such masses are almost completely convective. In 8 and 16 $ M_\odot$ helium stars the convective cores have radii of about 60 and 70 per cent, respectively, of the stellar radii, and occupy most of the stellar mass (Paczynski 1971). 

According to Zahn (1975, 1977) the tidal synchronization timescale for a star with a convective core and a radiative envelope is given by:
\begin{equation}
             1/t_{sync}  = 52^{5/3}\left(\frac{g_s}{R}\right)^{1/2}\frac{MR^2}{I} q^2 (1+q)^{5/6} E_2 \left(\frac{R}{a}\right)^{8.5} ,             
\end{equation}
where $q = M_2/M$ is the mass ratio of the companion $(M_2)$ and of the star to be synchronized $(M)$, and $g_s$, $R$ and $I$ are the surface gravity, radius and moment of intertia, respectively, of the latter star, a is the orbital radius and $E_2$ is the tidal torque constant for stars with a radiative envelope and a convective core. $E_2$ is proportional to $(R_c/R)^{6}$, where $R_c$ is the radius of the convective core (Zahn, 1975, 1977). Zahn (1975) calculated the values of $E_2$ for main-sequence stars of various masses. For such stars in the mass range 7 to 15 $M_\odot$ he found $E_2$ to be around $10^{-4}$. In order to correct for the much larger relative radius of the convective cores in helium stars, one has to multiply the $E_2$-values for main-sequence stars of similar masses with $(R_{cHe}/R_{cms})^{6}$ , where $R_{cms}$ is the relative radius if the convective core of the main-sequence star, and $R_{cHe}$ is the one of the helium star.  To this end we used for the 8$M_\odot$ helium star $(R_c= 0.7R)$ the $E_2$ value of Zahn's 10$M_\odot$ main-sequence star $(R_c = 0.27R)$ and for the 16~$M_\odot$ Helium star $(R_c= 0.8R)$ we used the $E_2$ value of Zahn`s 15$M_\odot$ main-sequence star $(R_c= 0.30R)$. This yields $E_2 = 4.4\times10^{-4}$ for the 8$M_\odot$ helium star and $E_2 = 1.7\times10^{-2}$ for the 16$M_\odot$ helium star.  

In order to get the shortest possible orbital periods, we now assume that after the CE phase the low-mass main-sequence companion of the helium star fills its Roche lobe.

We then find for the 8$M_\odot$ helium star that with Roche-lobe-filling companions of 1, 2 and 4 $M_\odot$, respectively, the orbital periods are 8.78, 10.45 and 12.43 hours, respectively; using equation (2) we then find that with these three companion masses the tidal synchronization timescales of these three systems are 1800, 1400 and 1130 years, respectively.
For a 16$M_\odot$ helium star the orbital periods with these three main-sequence companion masses are exactly the same and the tidal synchronization timescales are 440, 400 and 370 years, respectively. 
The lifetimes of helium stars of 8 and 16 $M_\odot$, respectively, are of order $5\times10^5$ yrs  (Paczynski 1971). Thus one expects, as already assumed by Izzard et al. (2004) and Podsiadlowski et al. (2004), that these helium stars will be fully synchronized with their orbital motion throughout their core-helium-burning evolution.
Could after the end of helium burning the contracting Carbon-Oxygen core of   the helium star keep the angular momentum which it obtained in its state of synchronized helium star and maintain that angular momentum until core collapse? As we will now show, it is unlikely that it will be able to take this barrier.

 \begin{figure}[ht!]
\begin{center}
\includegraphics[width=0.48\textwidth]{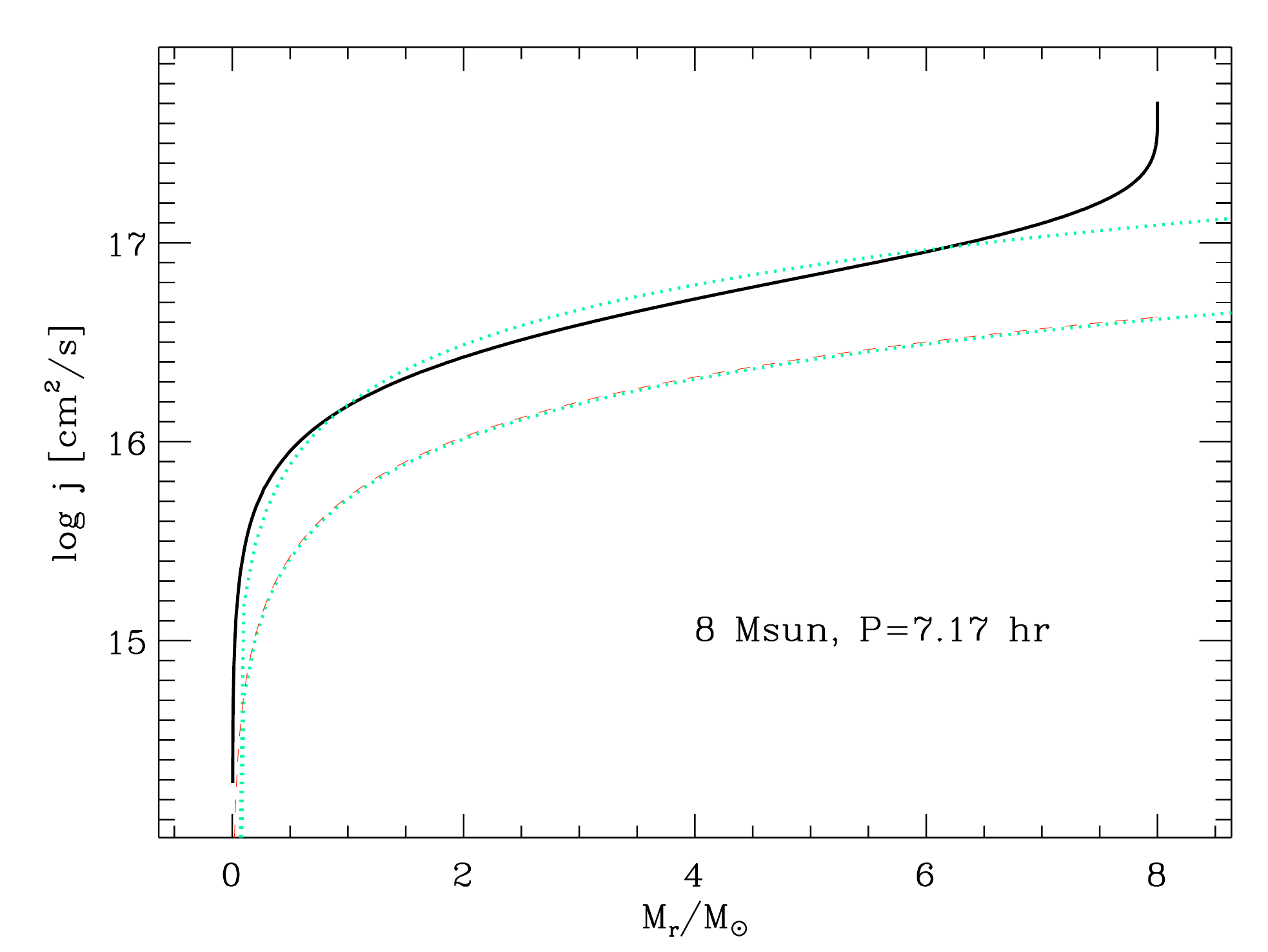}
\end{center}
\caption{Solid curve: specific angular momentum as a function of mass in a synchronized  8 solar mass helium star with a 0.8 solar mass Roche-lobe filling main-sequence companion. Dotted curve: specific angular momentum distribution required for the formation of a hypernova in case the mass interior to $M_r$ collapses to a Schwarzschild black hole; dash-dotted curve: the same for the case of a Kerr black hole }
\label{fig:1}       
\end{figure}

\subsection{ Timescales for core-envelope coupling}
\label{subsec:5.3}
 The fully drawn curve in Figure 2 shows the specific angular momentum distribution in a synchronized helium star of 8$M_\odot$ in a close binary with a 0.8$M_\odot$ Roche-lobe filling companion (Porb = 7.17h), compared with the minimum specific angular momentum required to form an accretion disk around a Schwarzschild and a Kerr black hole, as a function of the black hole mass. One observes from this figure that if the inner part of the helium star can maintain its specific angular momentum also when it becomes a contracting CO-core (which then will spin much faster than its helium envelope) then indeed the inner parts of such helium stars would be able to produce a hypernova/GRB if the black hole is of the Kerr type.
However, whether the contracting CO-core can maintain its specific angular momentum which it had as a helium star, depends on the timescale of core-envelope coupling. It is expected that this coupling in a convective differentially rotating star will be due to magnetic fields generated in this star, and Spruit (2002) has derived the order of magnitude timescale for this coupling. 
Yoon (2006) calculated the evolution of rotating helium stars with masses between 8 and 40 $M_\odot$ using Spruit's (2002) mechanism for core-envelope coupling. He found that the inner 3 $M_\odot$ of the CO-cores of these stars at the moment of core collapse have retained a fraction f of their initial specific angular momentum which they had as a helium star in solid-body rotation:
For $M_{He}$ = 8-16 $M_\odot$: f  = 0.2; 20$M_\odot$:  f  = 0.4;  25$M_\odot$:  f  = 0.6;  
30$M_\odot$: f  = 0.65;  40 $M_\odot$: f  = 0.75.

Using these values for 8-16 $M_\odot$ stars in Figure 2 one sees that the specific angular momentum in the central parts of the 8 $M_\odot$ helium star (the fully drawn curve) moves downwards by a factor 5 and thus falls below the Kerr as well as the Schwarzschild curves. The same holds for the 16$M_\odot$ helium star. This means that while a Helium star in a close binary with a Roche-lobe filling low-mass main-sequence star has achieved tidal synchronization during core-helium burning, still its core at the time of its collapse will be unable to produce a hypernova/LGRB. We thus see that the progenitors of the black holes in the Black-Hole X-ray Binaries with low-mass companion stars in all likelyhood did not produce a hypernova/LGRB.

\subsection{Timescales for synchronization of helium stars in close binaries with a compact  
       companion}
       \label{subsec:5.4}
       
       Such binaries will form by the spiral-in of a neutron-star or black-hole companion of a massive star in a wide High-Mass X-ray Binary (HMXB). Recently, with INTEGRAL such a wide system was discovered, consisting of a blue supergiant and a compact star in a 330 day orbit (Sidoli et al. 2006). [In HMXBs with orbital periods shorter than about 100 days, the compact star is expected to spiral into the core of its companion such that no binary will be left (e.g. Taam 1996)]. Presently three close X-ray binaries consisting of a helium star (Wolf-Rayet star) and a compact object are known: Cygnus X-3 $(P_{orb}= 4.8$ h; van Kerkwijk et al. 1992), and the extragalactic sources IC10 X-1 $(P_{orb} = 34.8$ h, Prestwich et al. 2007; ATel 955) and NGC 300 X-1 $(P_{orb} = 32.8$ h; Carpano et al. 2007).  
The shortest possible orbital periods of helium star plus compact star binaries will occur if the helium star fills its Roche lobe.  For helium stars of 8$M_\odot$ and 16$M_\odot$ these shortest possible orbital periods are 2.046 and 2.466 hours, respectively, independent of the mass of the compact companion. Using equation (2) one finds that the synchronization timescales in these systems are extremely short, of the order of years to decades at most, such that they will remain synchronized throughout their core-helium- burning evolution.
The specific angular momentum is here $3.7\times10^{17}$ and $6.0\times10^{17}$ cgs, respectively. As mentioned above, the cores of these stars can maintain some 20 per cent of this up till core collapse. Equation (1) shows that this is sufficient to make a LGRB/hypernova.
Thus the post-in-spiral remnants of HMXBs are suitable for producing Long GRBs.

Some example progenitor HMXBs that might produce a LGRB:
We use Webbink`s (1984) equation to calculate the ratio of the final and initial orbital radius in the case of Common-Envelope evolution (e.g. see also van den Heuvel 1994). We will assume that the product $\alpha\lambda$ = 1, where $\alpha$ is the efficiency parameter for the ejection of the envelope, and $\lambda$ is a parameter characterizing the density structure of the star.
Our first example is Cygnus X-1, for which we adopt a mass of 35$M_\odot$, with a 14$M_\odot$ helium core for the supergiant and a mass of 15$M_\odot$ for the black hole (e.g. Gies and Bolton, 1982, 1986). The initial orbital period of 5.6 days of this system then results into a final orbital period of 2.4 hours for the 14$M_\odot$ helium star plus the 15$M_\odot$ black hole. In this case the helium star will be very close to filling is Roche lobe, so we expect the final product of the Cygnus X-1 system to be able to produce a hypernova/LGRB when the core of the helium star collapses to a black hole.

A second example is the system of 4U 1223-62/Wray 977, which consists of a neutron star and a blue hypergiant (B1.5Ia0) in an eccentric orbit with P = 41.5 days (e.g. see Kaper et al. 2006). The hypergiant is likely to have a mass ~ 35$M_\odot$, so we again we assume here a helium core of 14$M_\odot$. For the neutron star we assume a mass of 1.8$M_\odot$  (like in the system of Vela X-1, which also is a very massive X-ray binary). Assuming the same values for alpha and lambda as in the first case, we find that the final orbital period after spiral-in is 2.1 hours, such that again the helium star just fits inside its Roche lobe. So also here the core of the helium star at the time of collapse will have enough angular momentum to make a hypernova/LGRB.

\section{Discussion and Conclusions}
\label{sec:concl}

We saw in section 4 that completely rotationally mixed single star evolution at relatively low metallicities ($Z \leq0.33$ solar) may well provide a viable model for the production of LGRBs/hypernovae. 
As to binary models: the results from section 5 show that, assuming Zahn's (1975, 1977) model for the tidal synchronization of helium stars in close binaries, massive helium stars with main-sequence companions will be quickly synchronized, within a few centuries to millennia, with their orbital motion. However, we find that as a consequence of efficient core-envelope coupling in the post-helium burning phase it is unlikely that these stars by the time of core collapse will have sufficient core angular momentum to produce a hypernova/GRB. 
On the other hand, if the companion of the helium star is a compact object and the helium star is close to filling its Roche lobe (implying a very short orbital period, of the order of a few hours) we find that by the time of core collapse the core can still have sufficient angular momentum to produce a hypernova/GRB.
The fact that we already know two potential progenitors of close helium star plus compact star companion binaries among the HMXBs within 3.5 kpc distance from the sun implies that there must be several dozens such progenitor systems in our galaxy. Assuming a lifetime of some 50000 years for the HMXB phase, and 25 such systems in the Galaxy, one would expect one hypernova/LGRB from such systems every 2000 years. This is about 5 per cent of the SN rate in our galaxy. Assuming that the GRBs are beamed within a cone of opening half-angle 5 degrees (Frail et al. 2001), we would expect to observe one LGRB from such binary systems per 2 million years from a Galaxy like our own.  

We note that although this binary model appears viable, it remains puzzling why LGRBs have such a strong preference for the small irregular starburst galaxies. A possible explanation might be that at low metallicity a much larger fraction of the massive stars collapses to black holes. 
In such galaxies one would already expect most of the persistent ``standard'' HMXBs (that is: the ones with massive blue supergiant donor stars) to harbour black holes, while then also the donor stars in such systems are likely to collapse to black holes. 
This would imply that, if indeed the LGRBs originate from binary systems, a considerable fraction of the hypernovae/LGRBs will be the formation events of close double black hole systems. 
We note that also Tutukov and Cherepaschuck (2004) have proposed that LGRBs are later evolutionary products of HMXBs. They assumed (but did not calculate) that the helium star plus compact star remnants from such systems would be synchronized and also assumed that the collapsing cores would have retained the angular momentum from the time a synchronized helium star. We have shown here quantitatively that this is indeed the case.

\begin{acknowledgements}
This research was supported in part by the National Science Foundation under Grant No. PHY99-07949. 
The first author thanks the Mount Stromlo Observatory for its hospitality during the conference and the Netherlands research School for Astronomy NOVA for financial support for participation in this meeting. 
\end{acknowledgements}\\\\

References:\\

\noindent
Bloom, J.S., Djorgovski, S.G., Kulkarni, S.R. and Frail, D.A., 1998, Ap.J. 507, L25-L28\\
Bloom, J.S. et al., 1999, Nature, 401, 253- 456.\\
Brown, G.E., Weingartner, J.C. and Wijers, R.A.M.J., 1996, Ap. J. 463, 297-304.\\
Carpano, S., Pollock, A.M.T., Prestwich, A., Crowther, P., Wilms, J., Yungelson, L. and Ehle, M., 2007, Astro-ph/0703270 (accepted as a Letter in Astron. \& Ap.).\\
Fryer, C.L. (editor), 2004, ``Stellar Collapse'', Kluwer Acad. Publishers, Dordrecht, 406 pp.\\
Fryer, C.L., 2006, New Astron. Rev. 50, 492.\\
Fryer, C.L. and Woosley, S.E., 1998, Ap.J. 502, L9-L12.\\
Frail, D.A. et al., 2001, Ap.J. L55.\\
Galama, T.J., Vreeswijk, P.M., et al., 1998, Nature 395, 670.\\
Galama, T.J. et al., 2000, Ap.J. 536, 185-194.\\
Fruchter, A.S. et al., 1999, Ap.J. 519, 13-16.\\
Fruchter, A.S., Levan, A.J., Strolger, L., Vreeswijk, P.M., Thorsett, S.E., Bersier,D., Burud, I, and 26 co-authors, 2006, Nature 441, 463.\\
Hadfield, L.J. and Crowther, P.A., 2006, MMRAS 368, 1822-1832.\\
Hjorth, J. et al., 2003, Nature 423, 847-850.\\
Israelian, G., Rebolo, R., Basri, G., Casares, J. And Martin. E.L., 1999, Nature 401, 142-144.\\
Iwamoto, K. et al. 1998, Nature 395, 672.\\
Izzard, R.G., Ramirez-Ruiz, E. And Tout, C.A., 2004, MNRAS 348, 1215.\\
Kaper, L., van der Meer, A., van Kerkwijk, M.H and van den Heuvel, E.P.J., 2006, Astron. Ap. 457, 595-610.\\
Kulkarni, S.R. 2006, talk presented at Kavli Institute for Theoretical Physics, March 2006.\\
Levan, A. et al., 2005, Ap.J. 624, 880-888.\\
Limongi, M. and Chieffi, A. 2006, Ap.J. 647, 483.\\
MacFadyen, A.I. and Woosley, S.E., 1999, Ap.J. 524, 262.\\
McClintock, J.E. and Remillard, R.A., 2006 in ``Compact Stellar X-ray Sources'' (editors W.H.G.Lewin and M. van der Klis), Cambridge Univ. Press, p. 157-213.\\
Mokiem, R., de Koter, A., et al. 2006, Astro-ph/0606403.\\
Nelemans, G. And van den Heuvel, E.P.J., 2001, Astron.Ap. 376, 950.\\
Paczynski, B. 1971, Acta Astron. 21, 1-14.\\
Podsiadlowski, P., Mazzali, P.A., Nomoto, K., Lazzati, D. And Cappellaro, E., 2004, Ap.J. 607, L17-L20.\\
Portegies Zwart, S.F. and McMillan, S.L.W. 2002, Ap.J. 576, 899.\\
Portegies Zwart, S.F., Baumgardt, H., Hut, P., Makino, J. And McMillan, S.L.W., 2004, Nature, 428, 724.\\
Portegies Zwart, S.F., Baumgardt, H., McMillan, S.L.W., Makino, J., Hut, P., and Ebisuzaki, T. 2006, Ap.J. 641, 319.\\
Prestwich, A. et al, 2007, ATel Nr. 955.\\
Sidoli, L., Paizis, A. and Mereghetti, S., 2006, Astro-ph/10890S.\\
Sokolov, V.V. et al., 2001, Astron. Ap. 372, 438-455.\\
Spruit, H.C., 2002, Astron. Ap. 331, 923-932.\\
Taam, R.E., 1996, in ``Compact Stars in Binaries'' (editors J. van Paradijs, E.P.J.van den Heuvel and E.Kuulkers), Kluwer Acad. Publishers, Dordrecht, p. 3-15.\\
Tutukov, A.V. and Cherepaschuk, A.M. 2004, Astronomy Reports 48(1), 39-44.\\
Van den Heuvel, E.P.J., 1994, in ``Interacting Binaries'' (eds. H.Nussbaumer and A.Orr), Springer, Heidelberg, p. 263ff.\\
Van den Heuvel, E.P.J. and Habets, G.M.H.J., 1984, Nature 309, 598-600.\\
Van Paradijs, J, Groot, P.J., Galama, T., Kouveliotou, C. et al., 1997, Nature 386, 686-689.\\
Vreeswijk, P.M. et al., 2001, Ap.J. 546, 672-680.\\
Webbink, R.F., 1984, Ap.J. 277, 355-360.\\
Van Kerkwijk, M.H., Charles, P.A., Geballe, T.R., King, D.L., et al. 1992, Nature 355, 703.\\
Wolf, C. and Podsiadlowski, P., 2006, \\
astro-ph/0606725v3\\
Woosley, S.E., 1993, Ap.J. 405, 273.\\
Woosley, S.E., Heger, A. and Weaver, T.A., 2002, Rev. Mod. Phys. 74, 1015.\\
Woosley, S.E. and Bloom, J.S. 2006, Ann. Rev. Astron. Ap. 44, 507-556.\\
Woosley, S.E. and Heger, A. 2006, Ap.J. 637, 914-921.  \\
 Yoon, S.-C. and Langer, N., 2005, Astron. Ap. 443, 643-648.\\
Yoon,S.-C., Langer, N. and Norman, C. 2006, Astron. Ap. 460, 199-208.\\
Zahn, J.P., 1975, Astron. Ap. 41, 329.\\
Zahn, J.P. 1977, Astron. Ap. 57, 383-394. \\

\end{document}